# Ultra-low threshold continuous-wave quantum dot mini-BIC lasers

Hancheng Zhong[1], Jiawei Yang[1], Zhengqing Ding[1], Lidan Zhou[1], Yingxin Chen[1], Ying Yu[1]*, Siyuan Yu[1]*

[1] State Key Laboratory of Optoelectronic Materials and Technologies, School of Electronics and Information Technology, Sun Yat-Sen University, Guangzhou 510275, China

**Highly compact lasers with ultra-low threshold and single-mode continuous wave (CW) operation have been a long sought-after component for photonic integrated circuits (PICs). Photonic bound states in the continuum (BICs), due to their excellent ability of trapping light and enhancing light-matter interaction, have been investigated in lasing configurations combining various BIC cavities and optical gain materials. However, the realization of BIC laser with a highly compact size and an ultra-low CW threshold has remained elusive. We demonstrate room temperature CW BIC lasers in the 1310 nm O-band wavelength range, by fabricating a miniaturized BIC cavity in an InAs/GaAs epitaxial quantum dot (QD) gain membrane. By enabling effective trapping of both light and carriers in all three dimensions, ultra-low threshold of 12 μW (0.052 kW/cm$^2$) is achieved. Single-mode lasing is also realized in cavities as small as only 5×5 unit-cells (~2.5×2.5 μm$^2$ cavity size) with a mode volume of 1.16(λ/n)$^3$. With its advantages in terms of a small footprint, ultralow power consumption, robustness of fabrication and adaptability for integration, the mini-BIC lasers offer a perspective light source for future PICs aimed at high-capacity optical communications, sensing and quantum information.**

Lasers with ultra-low threshold and compact size are highly desirable in photonic integrated circuits (PICs)[1-3], aiming at the application of optical communications[4, 5], chip-scale solid-state LIDAR[6], and quantum information[7, 8]. The general approach to realizing such lasers is to effectively trap light and boost light-matter interaction by embedding gain materials into few- or sub-wavelength scale optical cavities with high quality (Q) factor and/or small mode volume (V-mode)[2, 9, 10]. Among different types of cavities, a photonic crystal (PhC) slab consisting of periodic dielectric structures is a versatile platform to achieve high Q factor via introducing defect-type PhC modes[11-14] or photonic bound states in the continuum (BICs) modes[15-21]. The former achieves lateral confinements using distributed Bragg reflection and out-of-plane confinement based on total internal reflection and the latter are formed based on topological mechanisms of either symmetry protection or destructive interference (accidental BIC mode). For lasing action, the reported defect-type PhC lasers, while exhibiting extremely small V-mode and therefore ultra-low threshold[12, 22, 23], nevertheless suffers instability caused by sensitivity to the structural disorder[24]. In this regard, BIC lasers that may benefit from topological robustness[18] are one of the most promising alternative architectures. However, radiative BIC (quasi-BIC) modes in PhC slabs[25-31] or gratings[32] with high Q-factor are often realized requiring symmetry in the vertical (thickness) direction and extended lateral periodic structures to reduce in-plane light leakage, therefore intrinsically limiting their footprint to hundreds of unit-cells.

A further factor impeding the performance of BIC lasers is the poor carrier confinement and pumping efficiency. In contrast to defect cavity PhC lasers where the light is localized therefore effective carrier confinement can be achieved by burying the gain medium in the defect cavity only[33, 34], BIC lasers, with its modes diffusely distributed across the cavity, require optical gain (therefore carriers) distributed across the structure. The high surface-volume ratio results in high non-radiative recombination of carriers. Therefore, reported conventional BIC lasers displayed relatively low pumping efficiency, high lasing thresholds and were limited to operating under femto- or pico-second pulse pumping[25-32], imposing a great challenge in advancing photonic integration applications that requires highly compact and low threshold lasers.

Such challenges have recently been alleviated to some extent by merging the two BIC modes (super-BICs)[35] or combining the BIC mode with other mirror-like refection either by Fano-mirror[36], or by photonic heterostructure[37]. The super-BIC laser[35] demonstrated relatively low threshold, but still pulse-pumped, lasing in InGaAsP PhC slab with a footprint of (40×40) unit cells. The Fano BIC laser[36] demonstrated excellent coherent profile and a threshold of ~12 kW/cm$^2$ under continuous-wave (CW) pumping but required localizing the gain in the continuum region by utilizing semiconductor buried heterostructure to protect the spatial asymmetry of the Fano BIC mode. The BIC laser in photonic heterostructure[37] has scaled down the diameter of fundamental mode to ~30 unit-cells, however the challenge is from the instability of the monolayer transition metal dichalcogenide gain material. A scalable, CW operated, highly compact, and ultra-low threshold BIC laser has therefore remained elusive.

Very recently, a new kind of BIC mode termed as miniaturized BICs (mini-BICs)[38] was proposed, which combines a traditional BIC mode and a lateral photonic bandgap mirror in a cooperative way to trap light in all three dimensions, achieving a record high Q factor and rather small V-mode in silicon-based passive structures[38]. On the other hand, epitaxial quantum dot (QD) materials, due to their ability to three-dimensional confinement of carriers, have lower threshold[39], high temperature stability[40, 41], and in particular high tolerance to epitaxial defects or etching-induced surface defects[42-44], therefore could serve as an efficient gain material for BIC lasers by suppressing nonradiative recombination paths.

In this work, we present the realization of CW operated BIC lasers with low-thresholds and small V-modes by combining O-band InAs/GaAs epitaxial QD gain material with mini-BIC cavities. Benefitting from the three-dimensional confinement of both light and carriers provided by the mini-BIC cavity and the QD, we achieve CW single mode operation by tuning the lattice constant and the cavity size to match the cavity mode frequency to the heterostructure bandgap. CW lasing threshold as low as 12 μW (0.052 kW/cm$^2$) is achieved when the resonant wavelength just at the peak of QD material gain spectrum. Single-mode mini-BIC lasers with the cavity size down to 5×5 unit-cells (~2.5×2.5 μm$^2$) are also demonstrated, exhibiting a mode volume as low as 1.16($\lambda$/n)$^3$. These mini-BIC QD lasers with their small footprint, low power consumption and robustness in fabrication could contribute to the development of high density integrated light sources on photonic integrated circuit (PIC) chips.

**Device operational principle**

A schematic of our mini-BIC laser is shown in **Fig. 1a**. The mini-BIC cavity is fabricated in a three-layer InAs/GaAs QD stack with a density of $5.5 \times 10^{10}/cm^2$ per layer (the right panel of **Fig. 1a**) and a thickness $h$ of 556 nm. The PhC slab is embedded in the middle of an ultraviolet curing adhesive (Norland Optical Adhesive, NOA) with a refractive index of 1.54, to provide mirror-flip symmetry in the vertical direction. The in-plane cavity is formed by a PhC heterostructure, which consists of a square-lattice array of nanoholes (region A) surrounded by a boundary region (region B) with a transition region between them. **Fig. 1b** plots the calculated band diagrams of infinite PhC slabs that have the same lattice constants as regions A and B. Here the diameter of hole $r$ is 390 nm, and the lattice constant of regions A and B are set as $a$=495 nm and $b$=530 nm respectively. We choose the lowest-frequency fundamental TE mode in region A (TE A) as the lasing mode due to its larger feedback strength (coupling constant) and thus lower threshold than other higher-order modes[45, 46]. The frequency of TE A is tuned so that it falls within the gain spectral range of the O-band InAs/GaAs QDs with its peak in the vicinity of 1300 nm (See Supplementary Information Fig. S1).

To achieve effective light-trapping in the transverse direction, the energy of TE A, which is above the light cone of region A (shaded green region in **Fig. 1b**), is designed to be inside the bandgap of region B (TE B) that forbids lateral leakage (yellow region in **Fig. 1b**), so that region B acts as an almost perfectly reflective mirror. Different from the continuous bands in infinite PhC slabs, eigenstates of the confined PhC in region A are a series of discrete modes, as the continuous momentum space is quantized into isolated points with a spacing of $\delta k = \pi/L$, where $L = N_a \cdot a$ is the cavity length of region A. Therefore, each discrete mode can be represented by a pair of quantum numbers ($p$, $q$) or defined as $M_{pq}$, indicating that it is localized near ($p\pi/L$, $q\pi/L$) in the first quadrant of the momentum ($k$) space. **Fig. 1c** shows a typical distribution of four eigenmodes ($M_{11}$, $M_{12}/M_{21}$ and $M_{22}$) in the k-space, in which the modes of $M_{12}$ and $M_{21}$ are degenerate in energy due to the $C_4$ symmetry of the structure. Here the number of holes along the side of regions A and B is set as $N_a$=13 and $N_b$=10 respectively, and the gap between the two regions is fixed as $(a+b)/2$. The corresponding calculated mode magnetic field $H_z$ profiles of $M_{11}$, $M_{12}/M_{21}$ and $M_{22}$ are plotted in **Fig. 1d**, indicating V-modes of $7.46(\lambda/n)^3$, $7.43(\lambda/n)^3$ and $7.40(\lambda/n)^3$, respectively.

To further reduce cavity volume as well as search for single mode operation, we explore the effect of device size and lattice constant on the modes of the mini-BIC structure. We simulate a series of structures with $N_a$ varying from 15 to 5 and $a$ from 495 nm to 485 nm. As shown in **Fig. 1e**, the decrease of $N_a$ results in the increase of $\delta k$ (equal to $\pi/L$), moving all modes outwards away from the center $\Gamma$ point with increasing mode intervals. Only one mode, $M_{11}$, settles inside the bandgap of region B when $N_a$ deceases to 7, which indicates that single mode lasing may be realized in a smaller cavity with $N_a$ =7 or 5.

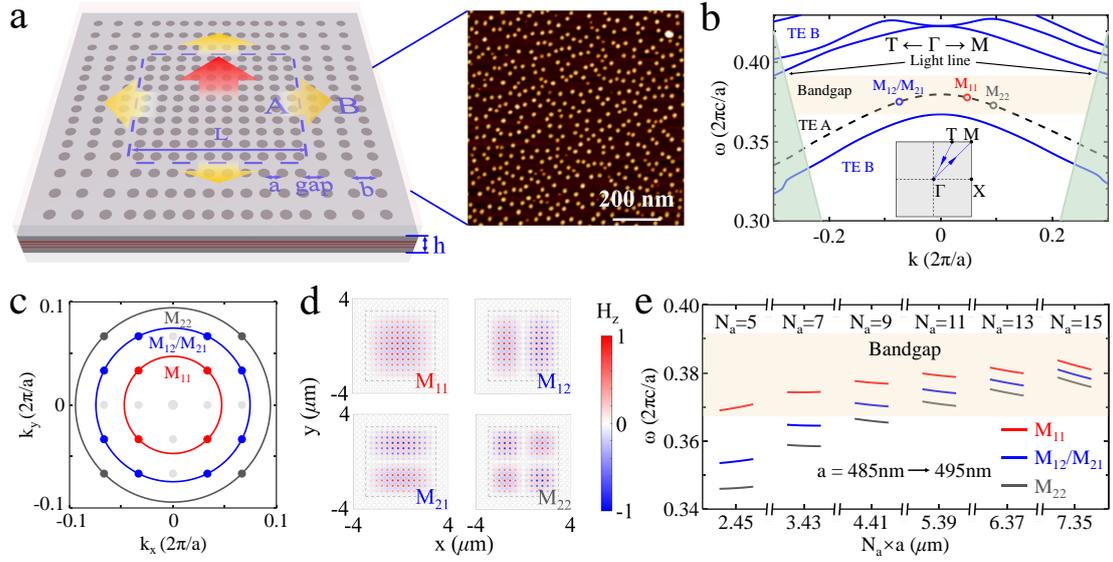

**Figure 1. The theoretical scheme of mini-BIC laser structure. a.** Schematic of a mini-BIC cavity (region A) encircled by a boundary of photonic bandgap (region B) to form a photonic heterostructure. The cavity region A is a $N_a \times N_a$ array of square latticed circular holes with the period of $a$ and side-length of $L$, while the boundary region B is an array with a width of $N_b$ and a period of $b$, but shares the same period of $a$ in the connecting side with region A. The circular holes of the PhC slab, whose radii equal to $r$, are etched in an InAs/GaAs QD active layer with thickness $h$ = 556 nm. The PhC slab is designed to be immersed in NOA. An atomic force microscopy (AFM) image of uncapped InAs/GaAs QDs (the right panel) indicates QD density of $5.5 \times 10^{10}$/cm$^2$. **b.** The calculated band diagrams of infinite PhC slabs: the continuous band (TE A, represented by the black dashed line) of an infinitely large PhC splits into a series of discrete modes above the light line and located in the bandgap of region B (TE B, represented by the blue solid line). **c.** The momentum distribution of each mode, labeled as $M_{pq}$ according to their location in the first quadrant of momentum space. **d.** Calculated $H_z$ near-field distributions of $M_{11}$, $M_{12}$, $M_{21}$ and $M_{22}$ at $a$ = 495 nm in a finite-size cavity with $N_a$ = 13. **e.** Simulated mode frequencies of mini-BIC structures with $Na$ varying from 15 to 5 and the lattice constant varying from 495 nm to 485 nm. The mode intervals increase with decreasing $N_a$ and only one mode of $M_{11}$ settles inside the bandgap when $Na$ deceases to 7, indicating that single mode lasing may be realized in a small cavity with $Na$ = 5 or 7.

## Lasing performance characterization

In experiment, the whole structure was fabricated on a GaAs-on-Glass platform using membrane transfer technique (details in Method section and Supplementary Information Fig. S2). To generate accidental BIC mode in experiment, the PhC membrane is placed in the middle of ~5 μm NOA ($n_{NOA}$ = 1.54) and sandwiched between two glass plates (n=1.49) to ensure the mirror-flip symmetry[17, 47]. **Fig. 2a** shows a typical top-view scanning electron microscopy (SEM) images of the fabricated mini-BIC cavity with $N_a$ =13. The photonic heterostructure of the mini-BIC laser can be clearly seen in the optical microscopy image as shown in the lower panel of **Fig. 2a**. The micro-photoluminescence (μ-PL) measurements are performed using a 705-nm CW laser with a spot size of ~5.4 μm at room temperature (details in Method section and Supplementary Information Fig. S3). **Fig. 2b** demonstrates the evolution of the emission spectrum under various pumping power of a mini-BIC laser with $a$ = 495 nm and $N_a$ = 13. A broad, spontaneous emission peak centered at 1303 nm is observed at low pumping intensities. With increasing pumping power, two sharp peaks appear at 1303 nm and 1316 nm (the resonant wavelengths of $M_{11}$ and $M_{12}/M_{21}$) and quickly dominates the emission spectrum with an overall suppression of the photoluminescence. At higher pump power, the weak peaks of other modes with larger quantum numbers (p and q) can also be excited, such as $M_{22}$ at 1326 nm and $M_{13}/M_{31}$ at 1333nm. It's worth noting that fabrication imperfections would slightly break the C4 symmetry and thus split the theoretically degenerate mode peaks of $M_{12}/M_{21}$ (**Fig. 2b**) with a minimal energy difference.

**Fig. 2c** shows the evolution of the output laser intensity (light in-light out (L-L) curve) and the linewidth of $M_{11}$ mode as a function of the pumping power, exhibiting a clear lasing behavior with a threshold power of 52 μW (0.227 kW/cm$^2$). The linewidth decreases from ~2.4 nm at low pumping power to 0.48 nm just below the threshold (the inset in **Fig. 2c**), which suggests a spectral linewidth narrowing effect during lasing and a cavity Q factor of 2715. The deviation between the simulated and measured Q factors (Supplementary Information Fig. S4) may be attributed to the vertical leakage and scattering loss caused by the inevitable fabrication imperfections. Nevertheless, due to the lateral light trapping by the photonic heterostructure, this measured Q factor is much larger than that from the structure without region B (Supplementary Information Fig. S5). Furthermore, the nearly constant lasing wavelength across the range of pumping power (less than 1.6 nm/mW in **Fig. 2d**) indicates that thermal effect was almost negligible in our devices, which may be attributed to the high temperature stability of QDs as well as the improved heat dissipation of our embedded BIC cavities compared with conventional suspended QD-PhC lasers[23], as the NOA has relatively higher thermal conductivity compared to air.

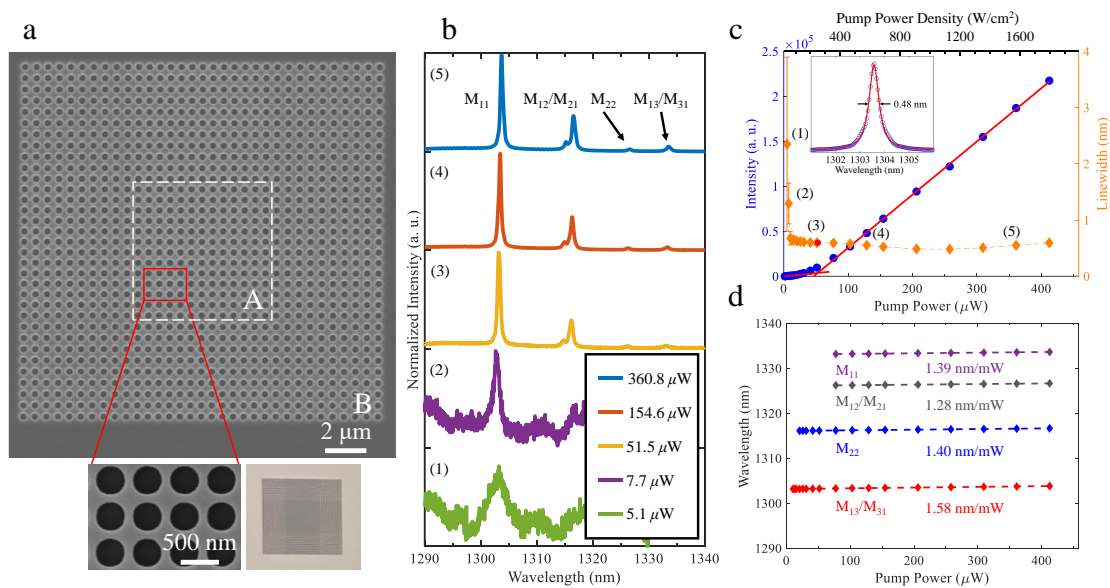

**Figure 2. Fabricated sample and the mini-BIC laser performance. a.** Scanning electron microscopy (SEM) images of the mini-BIC laser with $a$ = 495 nm and $N_a$ = 13, the inset illustrates a magnified view of the cavity region. The photonic heterostructure of the mini-BIC lasers can be clearly seen in the optical microscopy image (the lower

panel). **b.** Measured emission spectra under various pumping power. **c.** The collected emission intensity and the linewidth of the lasing $M_{11}$ peak at 1303 nm as a function of pumping power, indicating a lasing threshold of 52 μW (227 W/cm$^2$). The inset is a Lorentzian curve fitting of the spectra just below the threshold, which indicates a linewidth of ~0.48 nm and therefore a cavity Q factor of 2715. **d.** The lasing wavelengths of each mode under various input pumping powers (diamonds) and their linear fit (dashed lines). The error bars in c and d correspond to standard errors deduced by fitting.

## Single mode lasing with small footprint

To experimentally demonstrate single mode lasing and wavelength tunability, we vary the lattice constant $a$ from 495 nm to 485 nm in 2-nm steps and with $N_a$ changing from 5 to 15, resulting in a total of 36 devices that can be measured. The measured wavelengths of the all four modes agree well with the theoretical resonant wavelengths, which confirms that lasing action is indeed from the min-BIC discrete modes. As $N_a$ decreases, the wavelength of all modes shows a red-shift away from the wavelength at the $\varGamma$ point and simultaneously, the mode intervals increase due to the larger $\delta k$ (See Supplementary Information Fig. S6). Accordingly, fewer modes appear within the gain spectrum and thus fewer lasing modes are exhibited in the structure with smaller $N_a$. **Fig. 3a** further demonstrates the typical measured lasing wavelength of $M_{11}$ mode of mini-BIC cavities with different sizes and different lattice constants, where a wide tunable range near 80 nm is achieved, with highly predictable wavelengths. As expected, the experimental lasing wavelengths of $M_{11}$ red shift as $a$ increases due to the decreasing resonant frequency of TE A. The measured threshold powers of the lasers shown in **Fig. 3b** generally lie in the range of 12-75 μW except for the few far outlying in the long wavelengths and have a minimum close to the central wavelength of the ground state of the QD gain materials, where maximum gain is afforded by the strong zero-dimensional carrier confinement. Notably, an ultra-low threshold of 12 μW (0.052 kW/cm$^2$) is observed in the mini-BIC laser with $a$ = 487 nm and $N_a$ = 7, with a Q factor of 790 and a cavity mode exactly located at the peak (1300 nm) of the QD gain spectra (yellow region). A weak peak of $M_{12}/M_{21}$ mode at 1328 nm can be observed at high pump power (**Fig. 3c**), which may be attributed to the imperfect lateral optical confinement caused by insufficient number of periods ($N_b$=10) in region B. Single mode lasing is eventually achieved with the cavity size down to 5×5 unit-cells (~2.5×2.5 μm$^2$ with a mode volume of 1.16($\lambda$/n)$^3$). The L-L curve and lasing spectra of the device with $a$ = 485 nm are shown in **Fig. 3d,** exhibiting a threshold of 17 μW (0.074 kW/cm$^2$) and single-mode lasing across the range of pumping intensity up to 200 μW (12x threshold). More lasing spectra can be found in Supplementary Information Fig. S7.

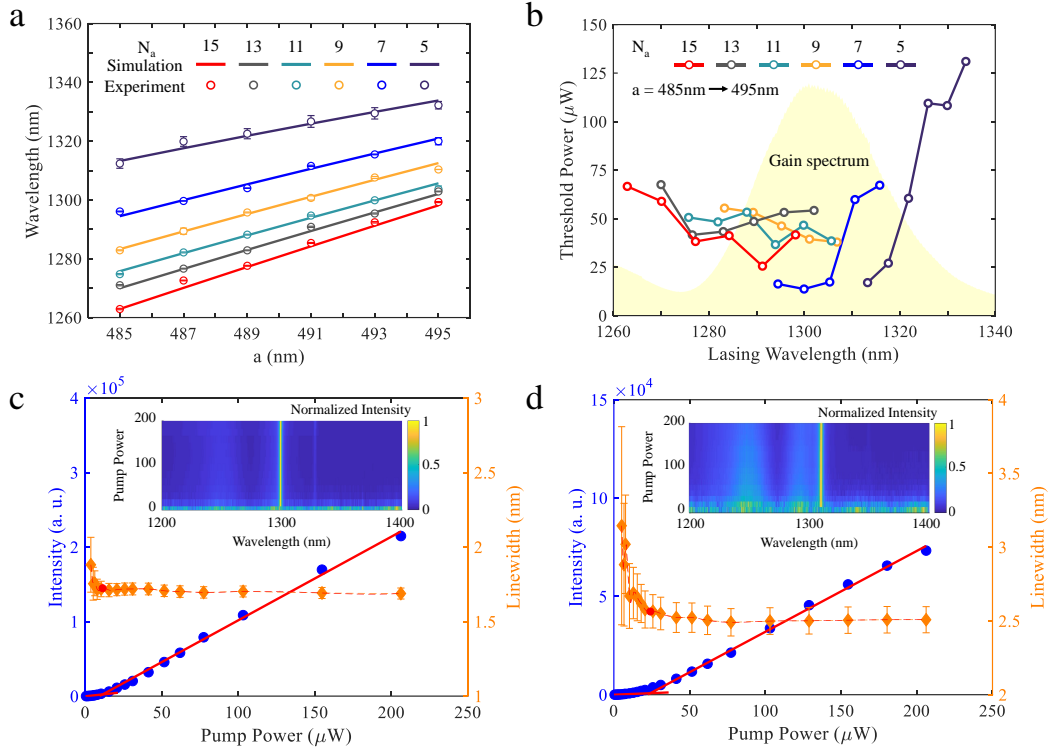

**Figure 3. Wavelength tunability and single mode lasing with small footprint. a.** The typical measured lasing wavelengths of $M_{11}$ mode in the cavities of different sizes and different periods agree well with theoretical resonant wavelengths. A wide tuning range of nearly 80 nm is achieved, with highly predictable wavelengths. **b.** The measured threshold power of the lasers with different $a$ and $N_a$. The yellow shaded region shows the relative magnitude of the QD gain spectrum. The threshold increases markedly for wavelengths longer than 1320 nm, where the optical gain of the QD ground state fall off sharply. On the short wavelength side, the threshold rises slowly as some gain results from the excited state of the QDs. **c-d.** The evolution of the collected emission intensity, linewidth as well as intensity spectrum as a function of pump power of the samples with $a = 487$ nm and $N_a = 7$ (c) and $a = 485$ nm and $N_a = 5$ (d).

## **Enhancement of the Q factor by topological engineering**

It is worth mentioning that the ultra-low threshold or single-mode operation data presented above represent some of the worst-case results in term of cavity Q factor, which on one hand confirms the robustness of the mini-BIC laser and on the other hand indicates that the mini-BIC lasers could be further improved to achieve even lower threshold.

Fundamentally, the Q factor could be further improved by the enhanced vertical light confinement, that is, the mini-BIC modes can be designed to converge with accidental BIC modes[18, 35, 38] by fine-tuning the lattice constant of region A. Taking the samples with a fixing $N_a$ of 13 and different lattice constant $a$ for example, as plotted in **Fig. 4a**, the high-Q ring arising from topological constellation shrinks towards Γ point as lattice constant $a$ increases and eventually reaches the states of $M_{11}$ at $a=495$ nm. Theoretically, extremely high Q factor of $\sim 10^{10}$ can be achieved in an 'empty' (i.e., no material loss) cavity due to this full three-dimensional light confinement. The theoretical trajectory of $M_{11}$ mode as a function of $N_a$ is labeled as circles along the Γ-M direction in the upper panel of **Fig. 4b**, where the momentum of $M_{11}$ with $N_a = 13$ should be nearest to the accidental BIC mode at $k = 0.057$. However, the measured Q factor, as shown in the lower panel of **Fig. 4b**, only shows an increase with the cavity size as the mode moves close to Γ point, without demonstrating the effect of the merge between the $M_{11}$ mode with the accidental BIC. We also track how the Q factor of $M_{11}$ changes with increasing lattice constant $a$ at varied $N_a$, and find it almost monotonically increases from $a = 485$ nm to $a = 495$ nm (**Fig. 4c**).

As mentioned, the experimental Q values in the range of 500-3000, which is far lower than above theoretical values of $Q_{intrinsic}$, are attributable to additional extrinsic losses caused by the vertical leakage and scattering losses due to the inevitable fabrication imperfections. For high Q cavities, the losses are additive and the Q factor can be expressed by $1/Q=1/Q_{intrinsic}+1/Q_{extrinsic}$, where $Q_{extrinsic}$, limited by the extrinsic losses, would increase with the surface-to-volume ratio of the cavity. It is likely that the low values of $Q_{extrinsic}$, which diminishes with the increasing duty cycle $r/a$, masks the high intrinsic Q factor expected of the merged constellation. The experimental Q factor can be further improved by increasing the period of region B or designing region B for each cavity configuration to ensure the lasing mode locates in the middle of its bandgap, so that lateral confinement is enhanced. Surface passivation that can smooth the etched hole sidewalls could also alleviate scattering losses.

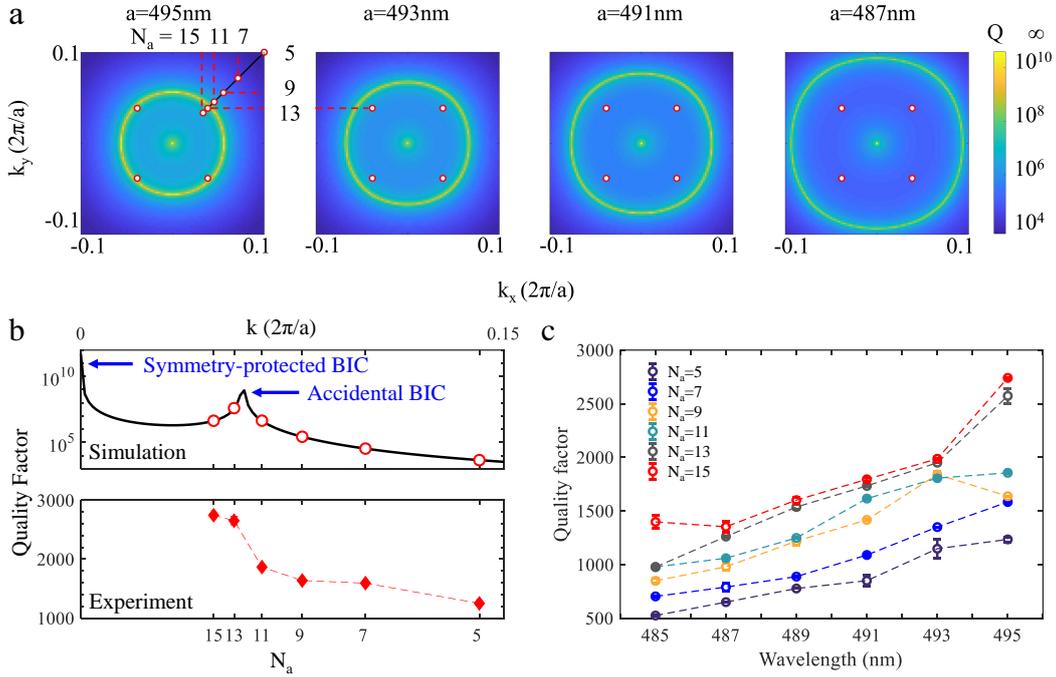

**Figure 4. Enhancement of the Q factor by topological engineering. a.** The high-Q ring arising from the constellation of multiple BICs appears on bulk band TE A in momentum space. When the period $a$ varies from 485 nm to 495 nm, the ring shrinks towards the center Γ point and eventually reaches the states of $M_{11}$ for $N_a = 13$ (red circles), increasing the Q factor of $M_{11}$ for $N_a = 13$. **b.** Simulated (red circles in the upper panel) and experimental (red diamonds connected by red dashed line in the lower panel) Q-factor of the lasing $M_{11}$ mode at $a = 495$ nm as the function of cavity size $N_a$ (lower axis) and therefore the function of wave vector k along Γ-M direction. Simulated Q factor of an infinite PhC is also plotted as black solid line, along which the symmetry-protected BIC at Γ point

and the accidental BIC at k = 0.057 can be seen. **c.** The measured Q factor of $M_{11}$ for varied $N_a$, which shows similar increasing trend as period $a$ increases.

**Conclusion**

Finally, we systematically compare the metrics of reported BIC lasers with different gain materials as well as those of our devices, as presented in **Table 1**.

In general, the generation of CW-pumped BIC lasers are boosted by combining BIC cavities with different lateral mirrors, such as Fano mirror[36] or PhC heterostructures[37], which significantly enhance the lateral confinement of light. Super-BIC cavity[35] can effectively improve the Q factor and therefore achieving relatively low-threshold lasing. However, lateral scattering loss and low pumping efficiency have so far only limited lasing to the pulsed mode. From the perspective of gain materials, quantum confinement materials, such as InGaAsP QWs[25, 28, 35, 36], monolayer $WS_2$[37] and Perovskite QDs[30], have achieved superior performance in lower threshold power due to their carrier confinement in several dimensions. Our devices, by implementing mini-BIC cavities in an InAs/GaAs QD gain material to achieve three-dimensional confinement of both light and carriers, have simultaneously achieved CW pumping with the lowest reported threshold power density and the smallest footprint.

To conclude, we successfully achieve CW pumped O-band mini-BIC lasers fabricated in an InAs/GaAs QD gain material. The smallest of the mini-BIC lasers has only 5×5 unit-cells with a small mode volume of $1.16(\lambda/n)^3$ and exhibits an ultra-low single mode lasing threshold of 17 μW (0.074 kW/cm$^2$), while the lowest threshold of 12 μW (0.052 kW/cm$^2$) is achieved in 7x7 unit cell devices. The lowest threshold power density is significantly reduced by as much as 99.6%, or down by more than two and half orders of magnitude, compared to the only reported monolithic CW BIC laser in III-V semiconductor QW gain material[36]. By careful engineering of structural parameters, the mini-BIC lasers have also been tuned across a wavelength range of 80 nm.

The mini-BIC lasers, fabricated by membrane transfer technique, can be flexibly implemented on different substrate such as silicon or $LiNbO_3$. Moreover, used as a surface-emitting laser based on transverse resonance, our mini-BIC lasers can have a noticeable advantage over vertical-cavity surface-emitting laser (VCSEL) at telecom/mid or far-infrared wavelength, which is based on vertical-cavity resonance and thus highly material dependent in terms of cavity construction. Just as photonic crystal surface-emitting lasers (PCSEL)[16] and topological-cavity surface-emitting laser (TCSEL)[48], the resonant wavelength of the mini-BIC lasers can be precisely tuned by simply varying the lattice constant of planar cavity, without the strict limitation imposed by the thickness of DBR material. On the other hand, the ability to engineer the lateral confinement can also lead to in-plane emission that couples directly into waveguides, thereby providing efficient, high spectral quality, precisely wavelength engineered miniature laser sources for PICs.

Moving forward, in addition to further optimizing cavity Q factor and reducing threshold, our approach can be combined with p-i-n hetero-junction structures to enable electrical pumped BIC-lasers, as its small size provides opportunities of reducing both optical and electrical losses during carrier injection and recombination. The ability to confine light and carrier in all-three dimensions may also open a pathway to strong light-matter interaction in BIC cavities leading to the realization of novel quantum light sources.

Table 1. Comparison of the mini-BIC laser with other BIC lasers.

| | BIC type | Pump Method | Gain medium | Wavelength (nm) | Cavity size($\mu m^2$) | Threshold peak power (mW) | Threshold power density ($kW/cm^2$) | Q factor | Ref. |
|---|---|---|---|---|---|---|---|---|---|
| BIC | Symmetry-protected & accidental BIC | Pulse pump | InGaAsP QWs | ~1600 | — | 73 | — | — | [25] |
| | | | InGaAsP QWs | 1551 | 19×19 | 15.6 | ~4 | ~4701 | [28] |
| | Symmetry-protected BIC | Pulse pump | GaAs | 830–850 | — | $8.8 \times 10^5$ | $7.0 \times 10^4$ | 2750 | [26] |
| | | | CdSe | 632-663 | — | $5.09 \times 10^8$ | $1.8 \times 10^5$ | 2590 | [31] |
| | | | Perovskite | 552 | — | $5.28 \times 10^5$ | $4.2 \times 10^4$ | — | [27] |
| | | | IR-792 molecules | ~860 | — | $\sim 2.16 \times 10^6$ | $\sim 2.75 \times 10^4$ | ~2883 | [29] |
| | | | Perovskite | 549 | — | — | $4.9 \times 10^5$ | 1119 | [32] |
| | | | Perovskite QDs | ~630 | — | — | 11 | — | [30] |
| BIC in PhC heterostructure | Symmetry-protected BIC | CW pump | monolayer $WS_2$ | 637 | 137 | — | 0.144 | 2500 | [37] |
| Super-BIC | Symmetry-protected & accidental BIC | Pulse pump | InGaAsP QWs | ~1600 | 23×23 | 0.34 | 1.47 | ~7300 | [35] |
| Fano-BIC | Fabry-Perot BIC | CW pump | InGaAsP QWs | 1560 | ~2.2 | 3.5 | 12.38 | ~78000 | [36] |
| Mini-BIC | Symmetry-protected & accidental BIC | CW pump | InAs/GaAs QDs | 1303/1328 (MM) | 3.4×3.4 | $1.2 \times 10^{-2}$ | 0.052 | 790 | This work |
| | | | | 1311 (SM) | 2.5×2.5 | $1.7 \times 10^{-2}$ | 0.074 | 525 | |

MM: multi-modes; SM: single mode.

**Method**

**Numerical Simulation.** The photonic band diagrams and mode characteristics are calculated using a three-dimensional finite-element method (FEM) solver of the COMSOL Multiphysics in the frequency domain. Three-dimensional models are built between two perfect-matching layers (PML), with Floquet periodic boundaries imposed on the four surfaces perpendicular to the slabs. The frequencies and quality (Q) factors of resonances can be obtained by the eigenvalue solver. The modal volume of a cavity is calculated referring to the formula: $V = \int \varepsilon(\mathbf{r})|E(\mathbf{r})|^2 d^3\mathbf{r} / \max[\varepsilon(\mathbf{r})|E(\mathbf{r})|^2]$, where $\varepsilon(\mathbf{r})$ is the material dielectric constant and $|E(\mathbf{r})|$ is the electric field strength[38].

**Growth.** The QD samples are grown on semi-insulating GaAs (001) substrates by a solid source molecular beam epitaxy (Veeco GENxplor system). A sketch of the heterostructure is shown in Supplementary Fig. S1(a). It consists of a 200 nm $Al_{0.8}Ga_{0.2}As$ sacrificial layer and a 556 nm GaAs layer. Three layers of high-density ($\sim 5.5 \times 10^{10}$ $cm^{-2}$) InAs QDs separated by 40 nm GaAs barriers are embedded in the middle of the GaAs layer. Each QD layer comprises 2.4 ML InAs covered with a 3.5 nm $In_{0.15}Ga_{0.85}As$ strain-reducing layer. Room-temperature photoluminescence (PL) emission peaking at 1300 nm was observed (Supplementary Fig. S1(b)) with a narrow full-width at half-maximum (FWHM) of 30 meV.

**Device fabrication.** We first fabricate the photonic crystal slab using electron beam lithography and dry etching processes. Then the top surface of the III-V wafer is bonded to a transparent quartz substrate with ~2.5 μm NOA61 via an ultraviolent curing process. Citric- and HF-acids are used to selectively remove the GaAs substrate and the $Al_{0.8}Ga_{0.2}As$ sacrificial layer. After the wet etching, the QD-containing PhC layer is then capped by ~2.5 μm NOA61 and a glass plate. The finished sample therefore has the PhC membrane placed in the middle of ~5 μm NOA ($n_{NOA}$ = 1.54) and sandwiched between two glass plates (n=1.49) to ensure mirror-flip symmetry. Full fabrication details are presented in the Supplementary Information.

**Optical characterization.** The sample is characterized by means of confocal micro-photoluminescence spectroscopy at room temperature. A 705-nm continuous-wave laser was used to optically excite the device via a 50x objective with a numerical aperture of 0.65. The spot size of the pump laser is ~5.4 μm (Supplementary Information Fig. S3). The emitted photons are collected by the same objective and sent to an InGaAs monochromator for spectrum characterization. The resolution of the spectrometer is ~0.20 nm.

**Acknowledgement**

This work is supported by the the National Natural Science Foundation of China (11704424, 62135012), the National Key R&D Program of China (2018YFB2200201), the Science and Technology Program of Guangzhou (202103030001), the National Key R&D Program of Guangdong Province (2020B0303020001), and the Local Innovative and Research Teams Project of Guangdong Pearl River Talents Program (2017BT01121).